# Main Magnetic Focus Ion Trap, new tool for trapping of highly charged ions

V. P. Ovsyannikov[a]

*Hochschulstr. 13, 01069, Dresden, Germany*

It is proposed to produce the highly charged ions in the local ion trap formed by a rippling electron beam in the focusing magnetic field. The experimental results demonstrate the presence of iridium ions with charges up to 50+. According to estimates, the average electron current density in the local ion trap can reach the value of the order of $1 \times 10^4$ A/cm$^2$. The pilot examples of devices of this type with the electron beam energies within the range 3-10 keV are also presented.

## I. INTRODUCTION

It is obvious that the choice of ion trap length substantially defines the design of Electron Beam Ion Source (EBIS) and problems in its construction. The first really working EBIS,[1,2] which was tested in Dubna in 1967, had the ion trap of 30 cm in length. It demonstrated possibilities of EBIS for production of highly charged ions. However, the following attempts to create EBIS with the length of ion trap of about 1 m and, correspondingly, with somewhat larger length of warm focusing solenoid turned out to be unsuccessful because of inhomogeneity of magnetic field.[3] It was the main reason for use of superconducting solenoids for focusing of extended dense electron beams in EBIS. The first superconducting EBIS (Krion 1) was designed in Dubna in 1972.[4,5]

Later on, the cryogenic technics and superconductivity became necessary attributes for the majority of EBIS around the world. Nevertheless, the use of this technics makes the devices of the CryEBIS type very complicated and expensive.[6] Apparently, it is connected with the difficulty of formation of the extended electron beam, with the diameter-to-length ratio of the order of about 1000.

The choice of ion trap of 2 cm in length made it possible to create the very successful electron beam ion trap (EBIT), the device for precision X-ray spectroscopy, where highly charged ions up to the bare uranium nuclei were observed. The choice of such length for ion trap is dictated by the danger of emergence of instabilities in dense electron beam and by the purposes of precision X-ray spectroscopy. In particular, the authors of work [7] remarked the following: "With this problem in mind, and because most of the instabilities are convective-like in nature, EBIT was designed to be as short as practical."

However, it is obvious that the length of ion trap can be significantly less than 2 cm on retention of a geometrical factor for efficiency of registration of radiation. It is achieved by reduction of the distance from the ion trap to detector in comparison to EBIT.

Therefore, we choose the ion trap of about 1 mm in length. Then the electron density in the ion trap, which is created by a rippling electron beam, is defined by focusing properties of a thick magnetic lens and can be significantly higher than it is predicted by the paraxial Brillouin/Hermann theory.[8] Consequently, in this case, the statement, that the Brillouin's electron current density in electron beam with the fixed current and voltage in a given magnetic field is the highest possible, becomes to be incorrect.

The local ion trap is created in the crossover of electron beam in thick magnetic lens, where under certain conditions the current density can reach colossal values.

## II. ELECTRON DENSITY IN LOCAL ION TRAP

As well known, the electron beam is focussed in the thick magnetic lens in the sequence of focuses, the first of which (or the main) is the most acute. The rippling electron beam creates a sequence of local ion traps according to the potential distribution along the source axis. Theoretically, the current density of the electron beam in the main magnetic focus can reach of enormous values.

Figure 1(a) presents the distribution of the electron beam current density in the first focus. From the Fig. 1(a), it is clear which phenomena will limit the achievement of very high density of electrons. These reasons are well known. Namely, there are aberrations of anode lens of the electron gun and the thermal velocity of the electrons. From the electron density distribution along the axis $z$ of the trap it can be distinguished two groups of electrons emitted from the central and peripheral parts of the cathode. Nevertheless, the average current density of electrons larger than $2 \times 10^4$ A/cm$^2$ on the length of about 1 mm can be obtained. For this purpose, the special forms of electron gun and of the magnetic field distribution have been found (Fig. 1(b)).

In Fig. 1(b), the electron trajectories are also shown along the source axis on the length of 1 mm for the electron beam of 50 mA at the energy of 10 keV in a short

---

[a]Author to whom correspondence should be addressed. Electronic mail: v.ovsyannikov@yandex.ru

magnetic field. The electrons are emitted from a flat cathode with diameter of 0.5 mm.

Additionally, in the classical theory of electron beam focusing by the thick magnetic lens, there is a fundamental effect of ion compression of the electron beam, which was formed from cathode in the zero magnetic field. The electron density should increase in the main magnetic focus with increasing the ion compensation in local ion trap. For the first time, this phenomenon was observed by J Arianer [9] and referred to as "supercompression".

Thus, the problem of focusing of the electron beam in Main Magnetic Focus Ion Trap (MaMFIT) is the opposite to the Brillouin focusing. In this case, the purpose of the focusing is to obtain the highest density of electrons in the focuses of the magnetic lens, in contrast to the equilibrium Brillouin flow. This problem, especially with taking into account of the ion focusing in the local ion trap, requires a special consideration and it will be published elsewhere.

### III. IONIZATION IN MaMFIT

For all computer simulations, the original code EBIS_T written by I Kalagin [10, 11] has been used. The results of computer simulations for ionization of ions in the trap with electron density of $2 \times 10^4$ A/cm$^2$ are shown in Figs. 2 and 3. Calculations are carried out for two ranges of electron energy of 3–10 keV and of 60–200 keV, respectively. For relatively light working gases up to Ar, the pressure in the ion trap is a constant, while the residual gas is the same as the working gas. For very highly charged $Xe^{52+}$ and $Pb^{80+}$ ions, the classical procedure of the "evaporative cooling" with He ions was used.[7]

### IV. PILOT EXAMPLES OF MaMFIT

The pilot examples of MaMFIT are based on the original design. This design allows one to construct very compact devices. Two rings, consisting of the rectangular permanent magnets, create the focusing magnetic field. The magnetization vector for each coil has a radial direction. These two coils with the opposite directions of magnetization are joined together by the external iron jacket and compose the complete magnetic focusing system. Finally, magnetic field of the solenoid type can be achieved in the space between the coils.

The magnetic field distribution is designed for obtaining a few magnetic focuses. A peculiarity of the design is the short gap between the magnetic coils for electron beam with energy up to 10 keV. The length of rippling $\lambda_r$ is estimated according to the formula

$$\lambda_r (mm) = 212 \frac{\sqrt{U}}{B} \sqrt{\frac{2}{1+K}},$$

where $U$ is the voltage of the electron beam (in V), $B$ is magnetic field (in Gs) and $K$ is cathode condition ($K = 0$ for Brillouin flow and $K = 1$ for fully immerse gun). This length is only of 4 mm for the Brillouin electron beam with the energy of 3 keV in magnetic field of 4 kGs.

Therefore, two types of MaMFIT have been developed. These very small devices of the first type (MaMFIT-3 and MaMFIT-10) for the energies of 1-10 keV are shown in Figs. 4 and 5. The devices of the second type have been designed for the electron beams with high energies of 60 keV and 200 keV (see Fig. 6).

### V. FIRST TESTING RUN OF MaMFIT-10

The first testing run of the ion source MaMFIT-10 was carried out in the Institute of Atomic and Molecular Physics in Giessen (Germany). The source was installed on the standard vacuum cross CF35 with a turbopump and a vacuum gauge as shown in Fig. 7.

The vacuum was measured at the exit of the source, directly above the pump. During the experiments the vacuum was not better than $3 \times 10^{-8}$ mbar with the current electron beam in direct circuit (DC) mode. Nevertheless, the device demonstrated the fine quality of the electron – optic system. The total intercepted current was only 20 – 40 μA for the electron current of 40 – 45 mA within the energy range of 7 – 8.5 keV.

The spectra of X-ray radiation from the iridium ions (cathode material) are presented in Fig. 8. The detector resolution is within the range of 200-240 eV. The charge states of Ir ions were identified by definition of positions of the radiative recombination picks in X-ray spectra and their comparison with the theoretical values for ionization energies.[12] As can be seen, the largest charge states of Ir ions are 49+ and 50+. The most of ions have the charge of about 44+.

The results of theoretical calculations are shown in Fig. 9. For the computer simulation of the experimental results, the codes [10,11] for electron density of $1 \times 10^4$ A/cm$^2$ and $2 \times 10^4$ A/cm$^2$ were used under some assumptions. First, it was assumed, that the molecules of iridium, which are evaporated from the cathode, create the permanent pressure of about $5 \times 10^{-10}$ mbar. Second, the background pressure is mainly due to hydrogen. The background pressure in the trap, which is calculated by using the vacuum equation, vacuum conductivity and the vacuum pressure above the pump, cannot be better than $3 \times 10^{-7}$ mbar.

The pick of charge state distribution is $Ir^{42+}$ for $1 \times 10^4$ A/cm$^2$ and $Ir^{46+}$ for $2 \times 10^4$ A/cm$^2$. Therefore, one can argue that the electron current density of the order of $1 \times 10^4$ A/cm$^2$ has been achieved.

This is a preliminary conclusion, which will be tested in future experiments. The computer simulations show that the iridium ions can reach the charge states around $Ir^{58+}$ after improvement of vacuum by a factor of two.

The peculiarities of physical processes in this unique ion trap will be discussed in details in following publications. The spectra given in Fig. 9 are described by the theoretical model with the relative low permanent pressure of the working substance and high permanent concentration of background gas (basically hydrogen).

## VII. CONCLUSIONS

Concluding, we have proposed to produce highly charged ions in the main local ion trap, which is formed by rippling electron beam in the focusing magnetic field. Based on this idea, a new generation of ion traps MaMFIT is developed. The electron current density of the order of $1 \times 10^4$ A/cm$^2$ is achieved in the local trap. The device can be employed for the precision X-ray spectroscopy and for extreme ultra violet research. After testing of original systems of the ion extraction, MaMFIT will be transformed into the ion source (MaMFIS). In this case the device can be used for investigation of interaction of highly charged ions with the solid-state surface and of single-ion implantation.

## ACKNOWLEDGEMENTS

The author wants to great thank A. Müller for giving the opportunity to perform the experimental research of the devices. The author is also grateful to A. Borovik, and K. Huber for their support in the test experiments and X-ray measurements. Aleksandr A. Levin, A. Nefiodov and A. Gorbunoff are acknowledged for their support. The author expresses also deep gratitude to O. K. Kultashev for his contribution to creation of the electronic optics.

**FIGURES:**

FIG.1. Electron current density distribution (a), magnetic field distribution $B(z)$ (b). Inset in (b) exhibits electron trajectories around the first focus position.

FIG. 2. Theoretical spectra of the Ne$^{i+}$ (a) and Ar$^{i+}$ (b) ions. Permanent pressure of the working gas is of about $1 \times 10^{-7}$ Torr.

Fig. 3. Theoretical spectra of Xe$^{i+}$ (a) and Pb$^{i+}$ (b) ions assuming the pulse injection of working gas. The permanent pressure of the cooling He gas is equal to $1.5 \times 10^{-9}$ Torr.

FIG. 4. General design (a) and overview (b) of MaMFIT-3 ($I_e$ = 50–100 mA, $E_e$ = 1–3 keV).

FIG. 5. General design (a) and overview (b) of MaMFIT-10 ($I_e$ = 50–100 mA, $E_e$ = 3–10 keV).

FIG. 6. Proposal design of MaMFIT with the electron energies of 60 keV (a) and 200 keV (b), respectively.

Fig. 7. Overview of MaMFIT-10 over the vacuum stand.

FIG. 8. X-ray spectra from iridium ions prepared by the electron beam with the energies of 7 keV (a) and 8.5 keV (b), respectively. The measurements were performed in the Institute of Atomic and Molecular Physics (Giessen).

FIG. 9. Theoretical charge state distribution of iridium ions in local ion trap for electron density $1 \times 10^4$ A/cm$^2$ (a) and $2 \times 10^4$ A/cm$^2$ (b), respectively.

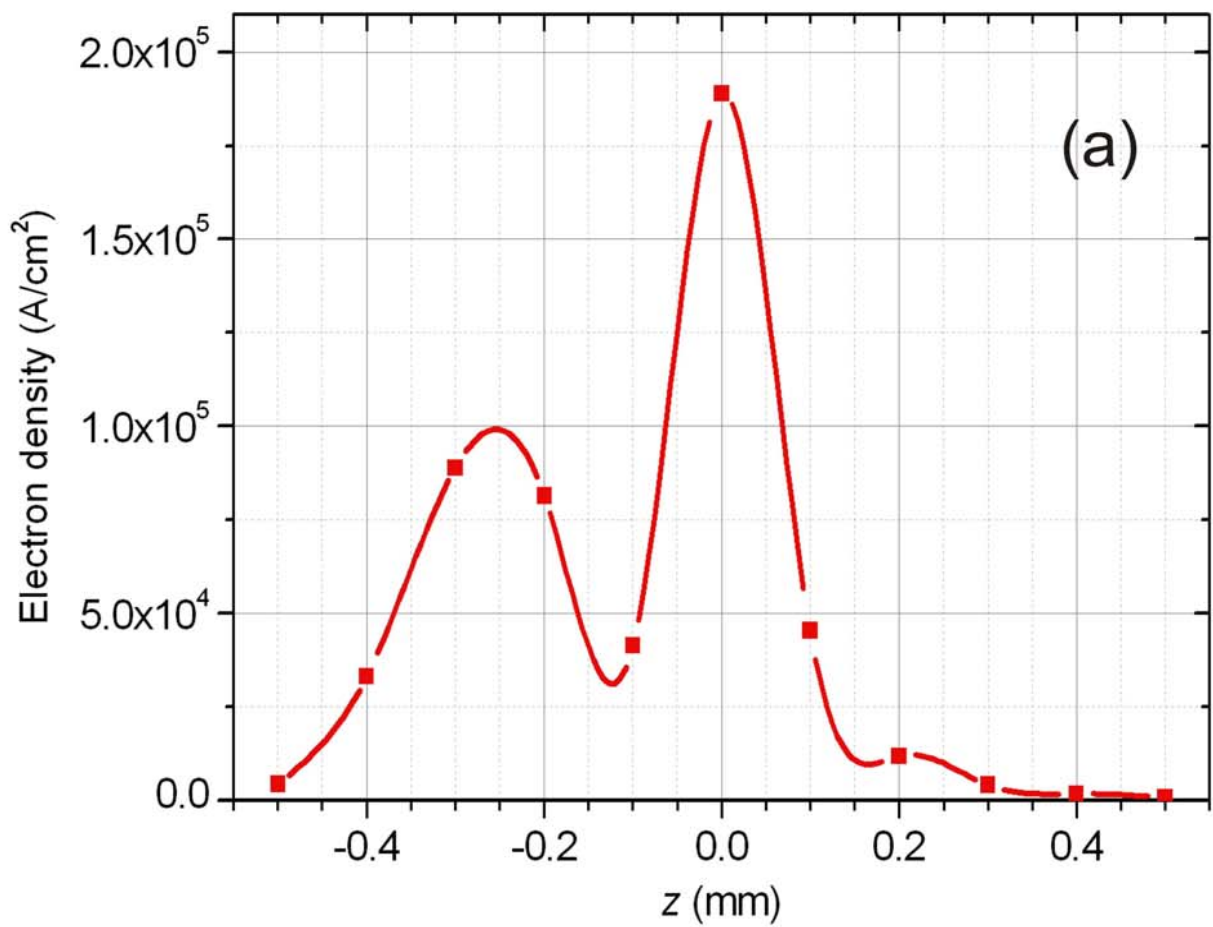
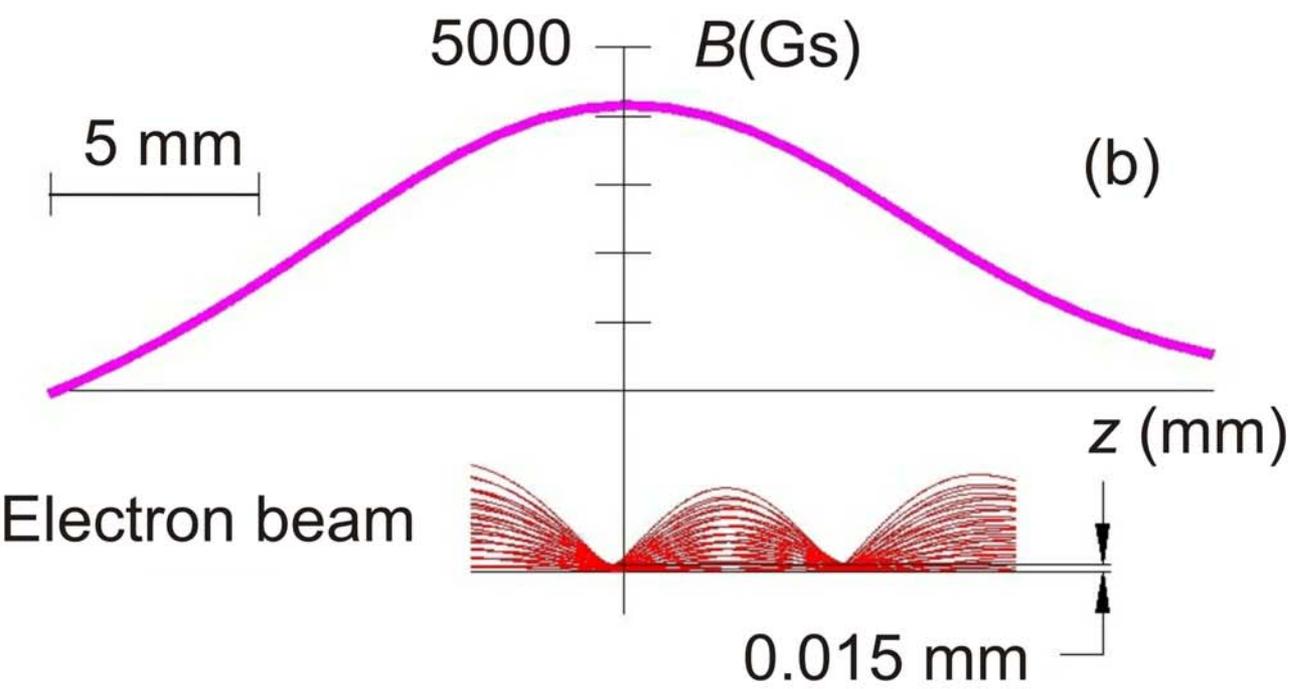

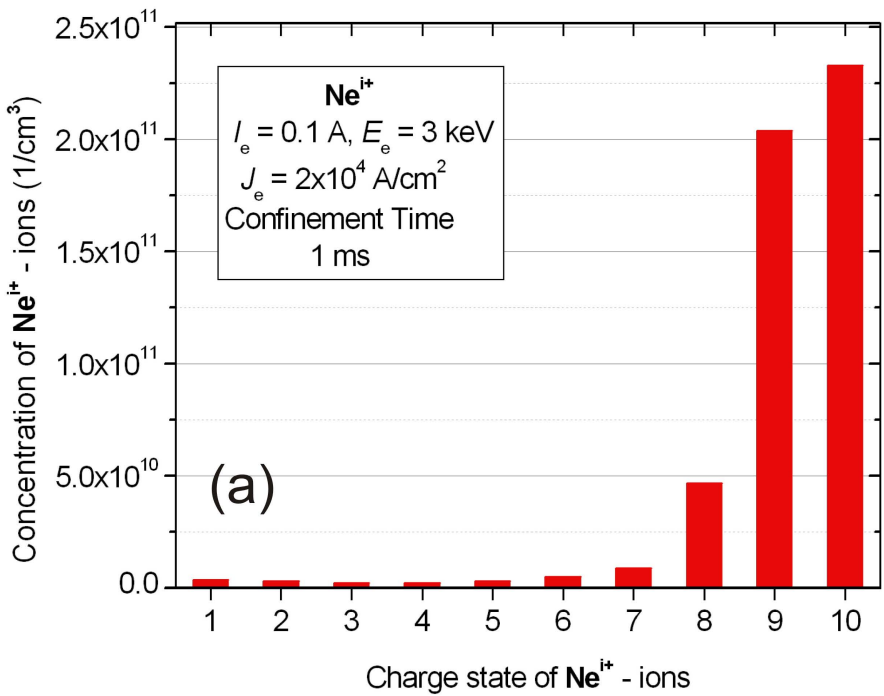

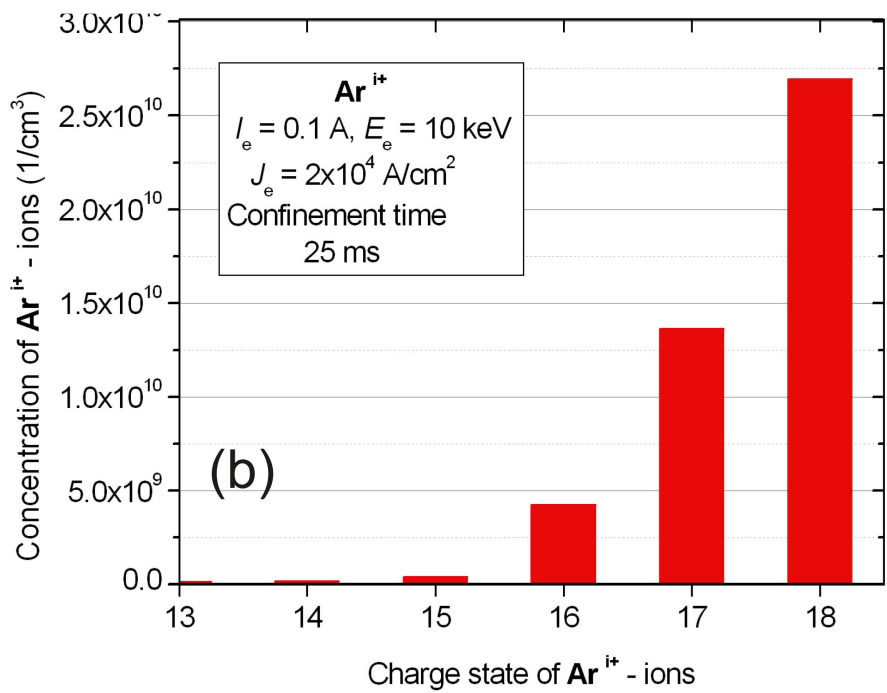

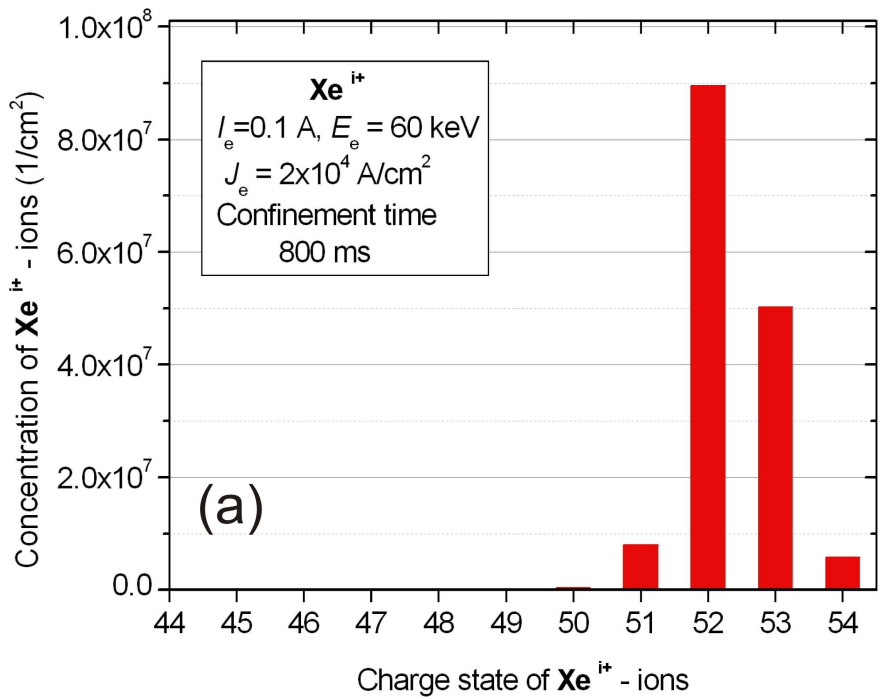

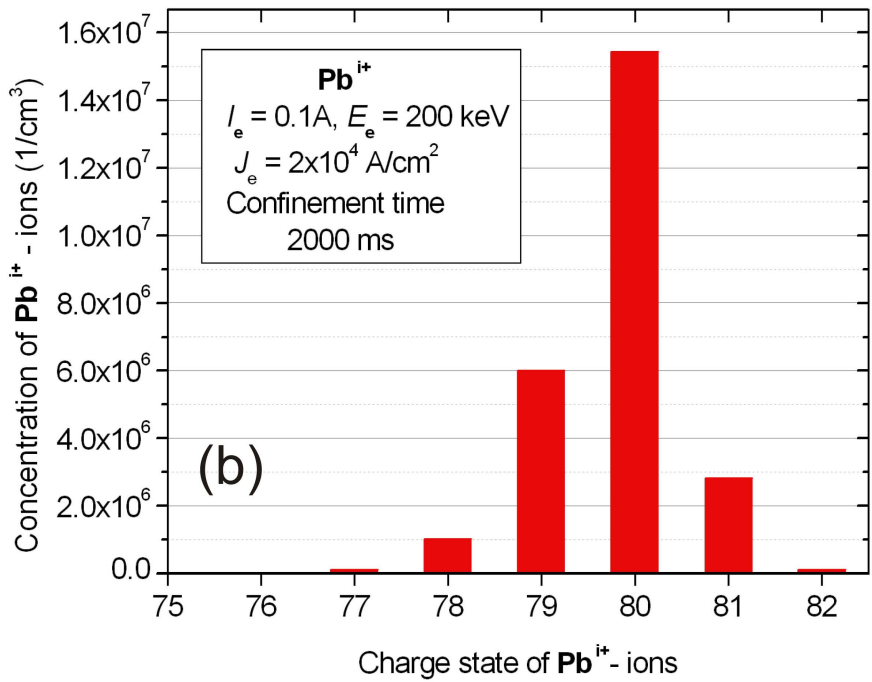

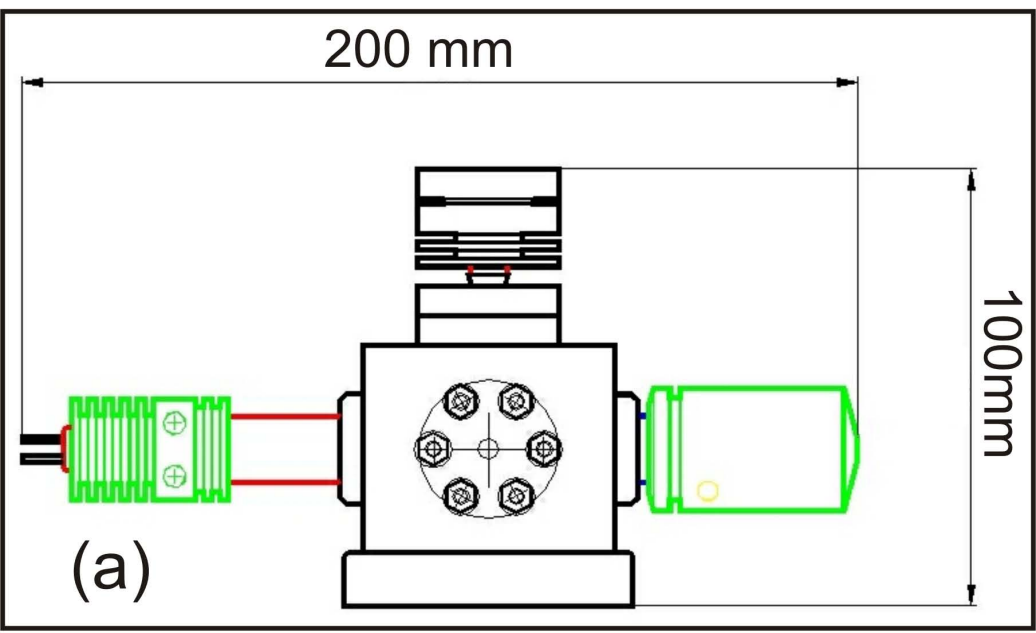
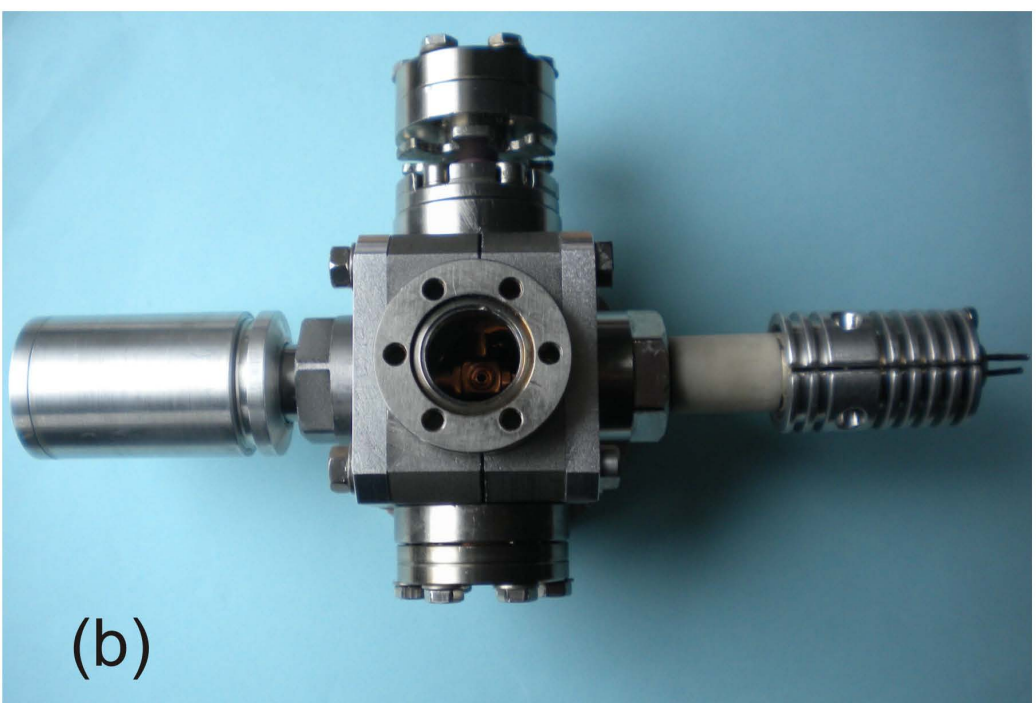

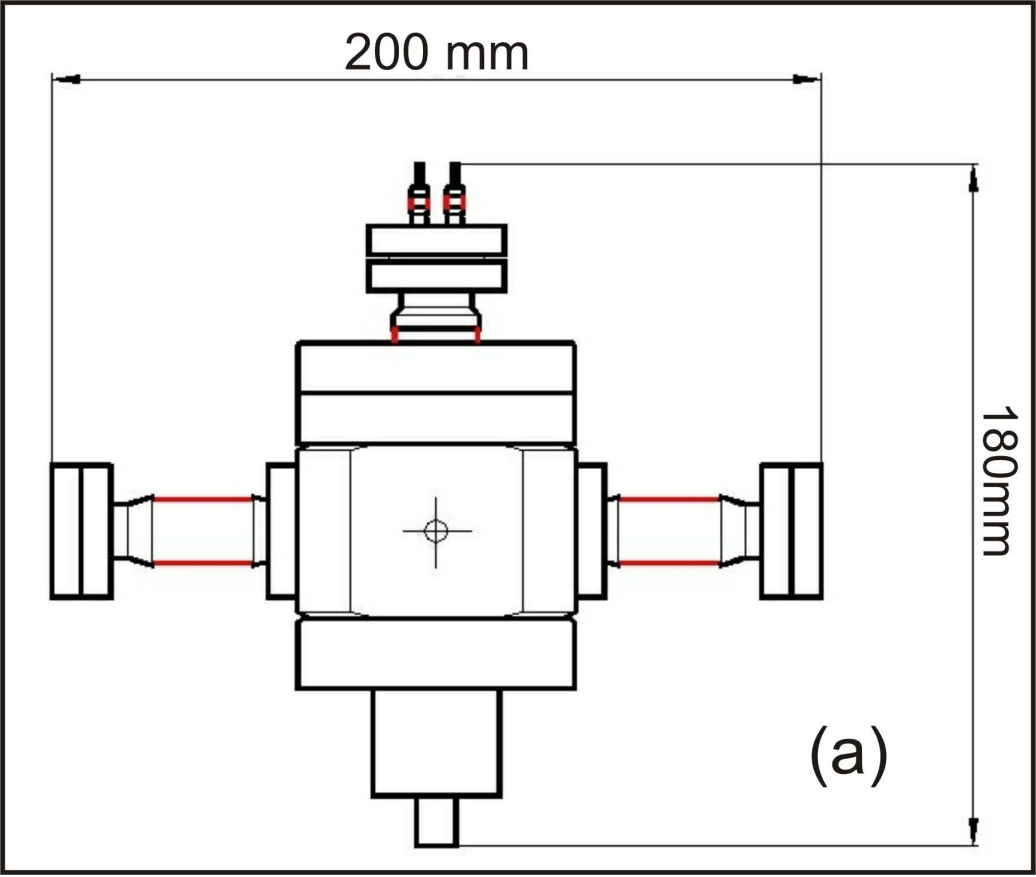
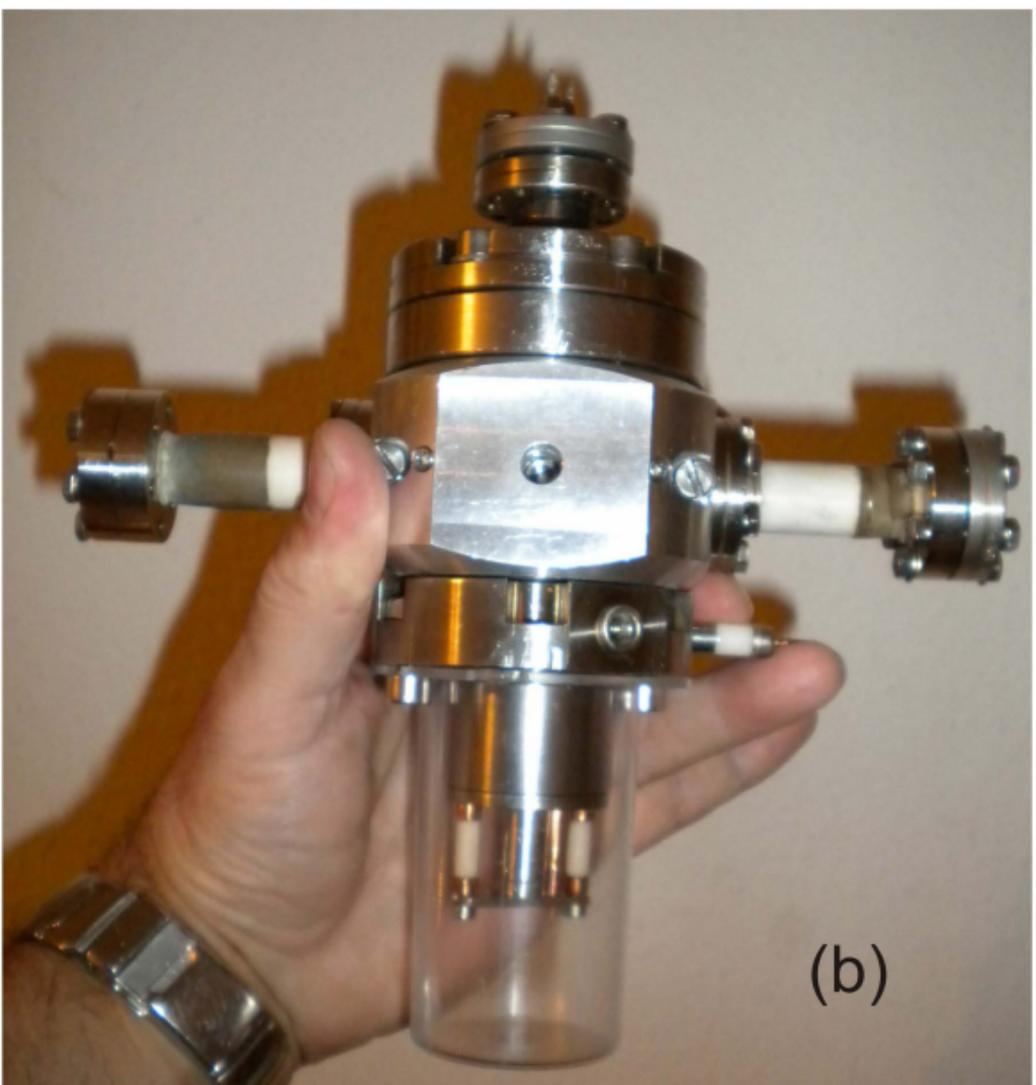

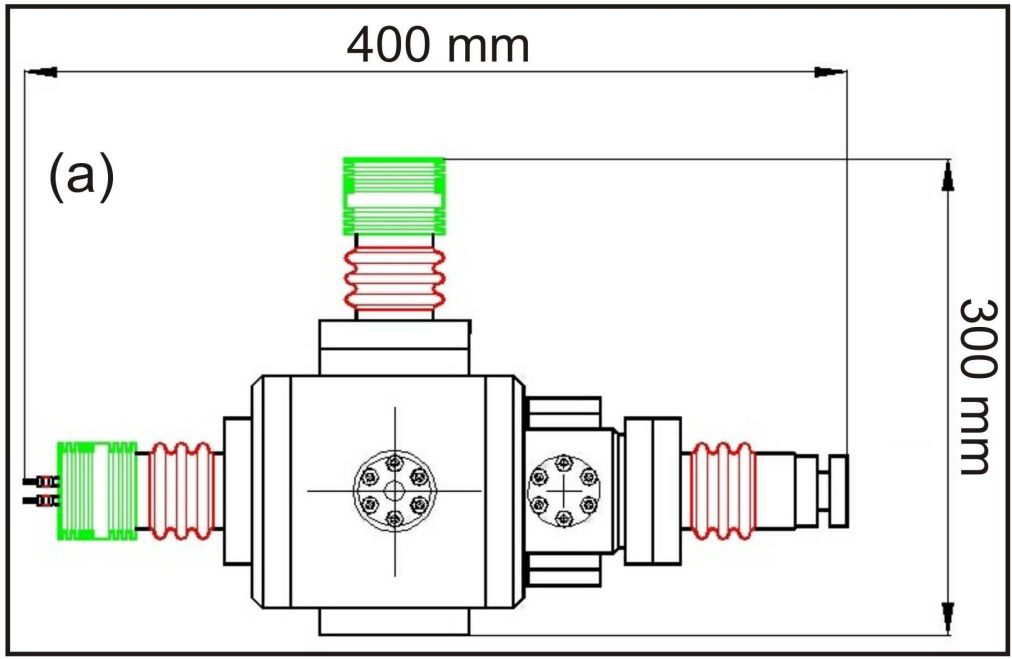

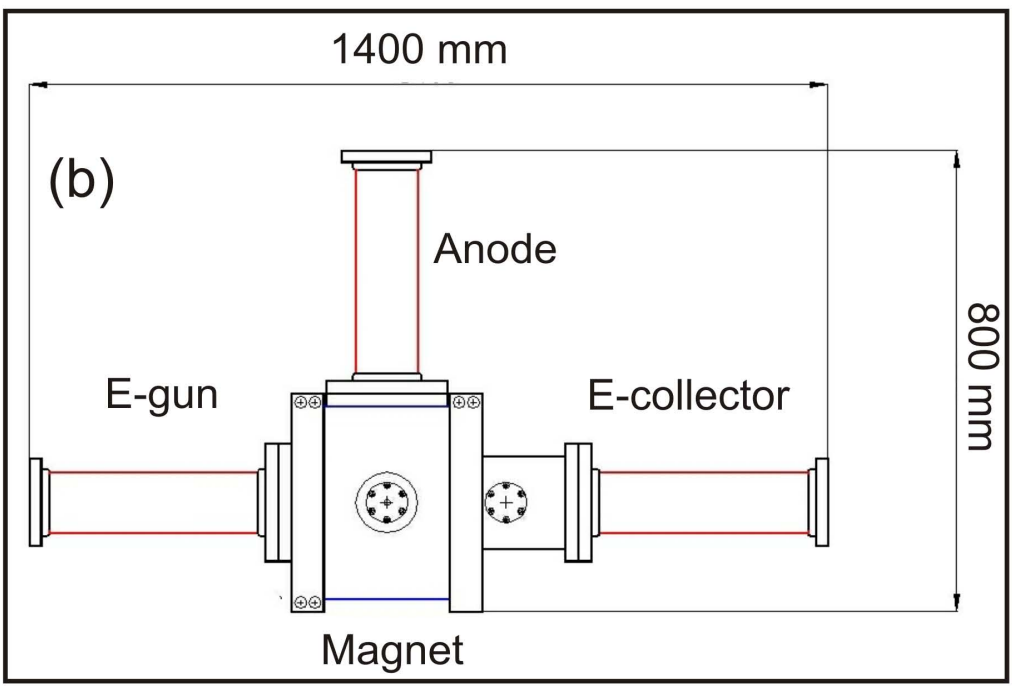

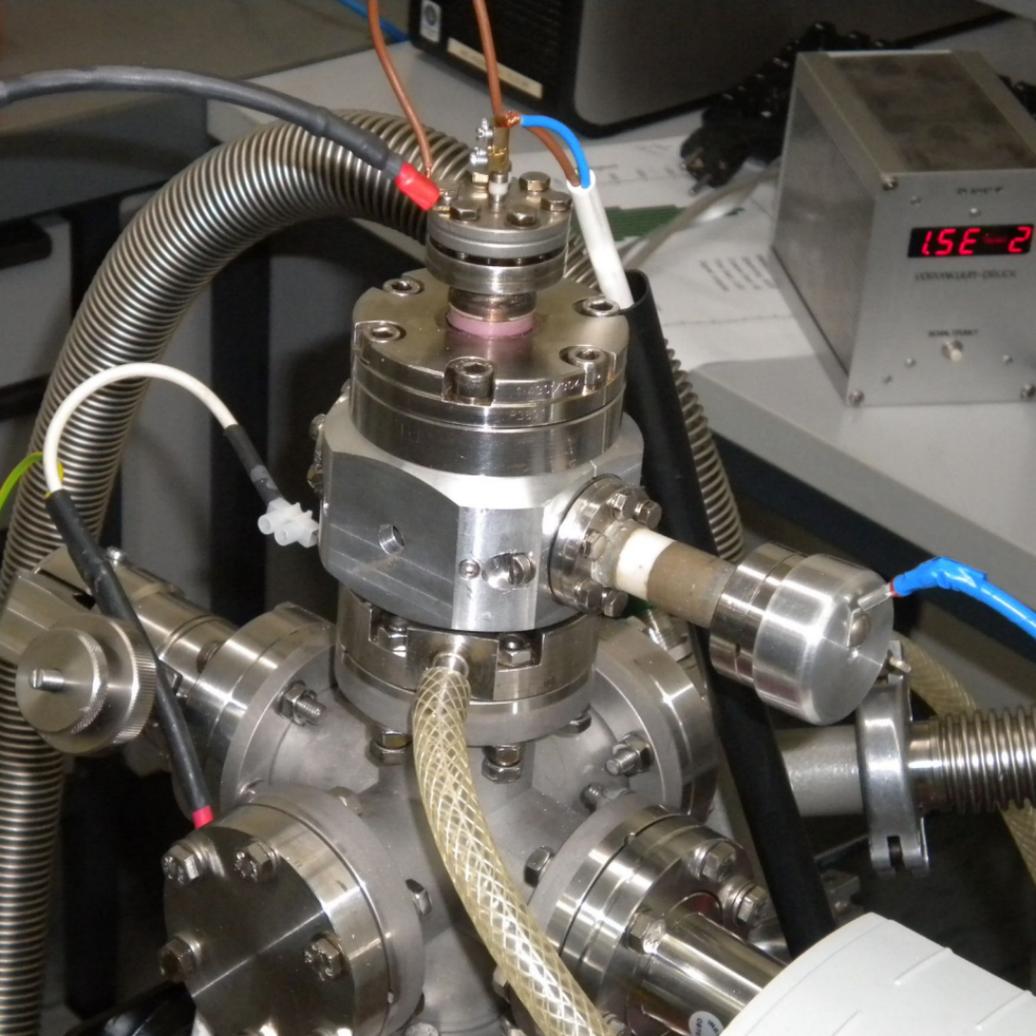

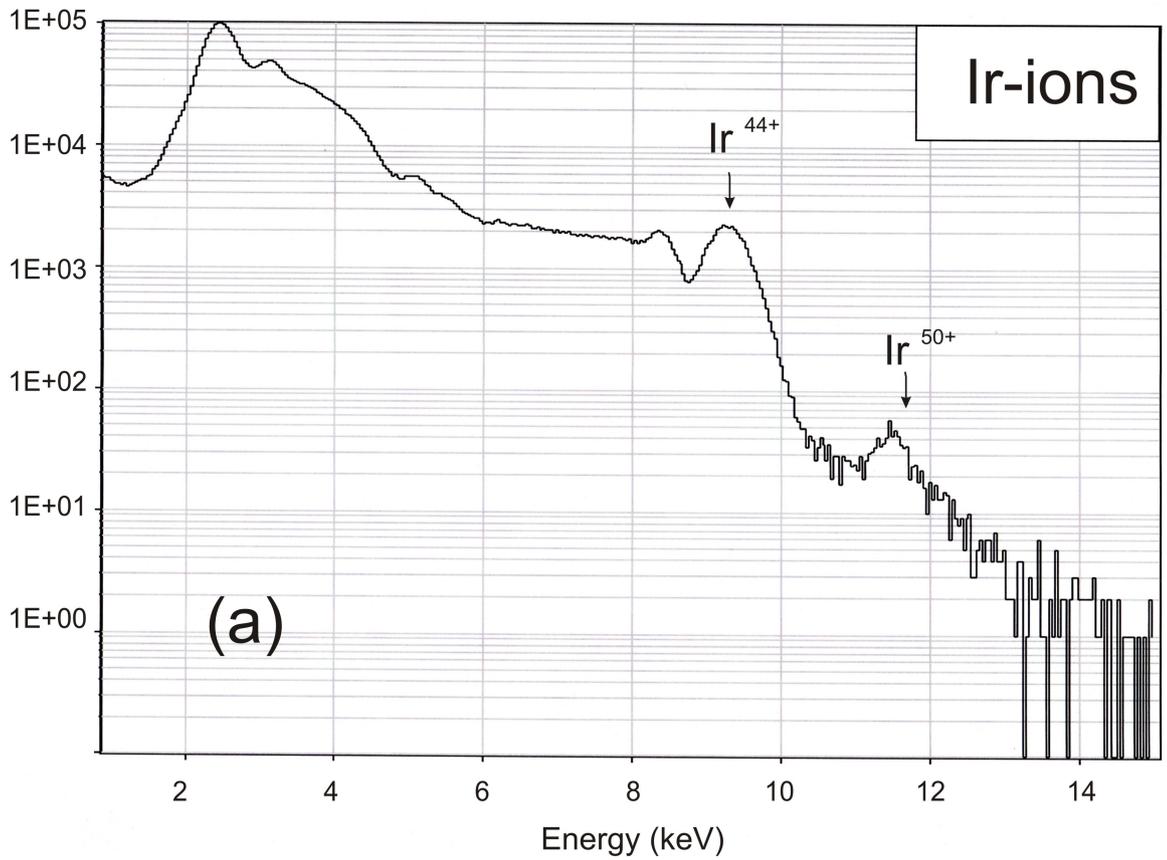

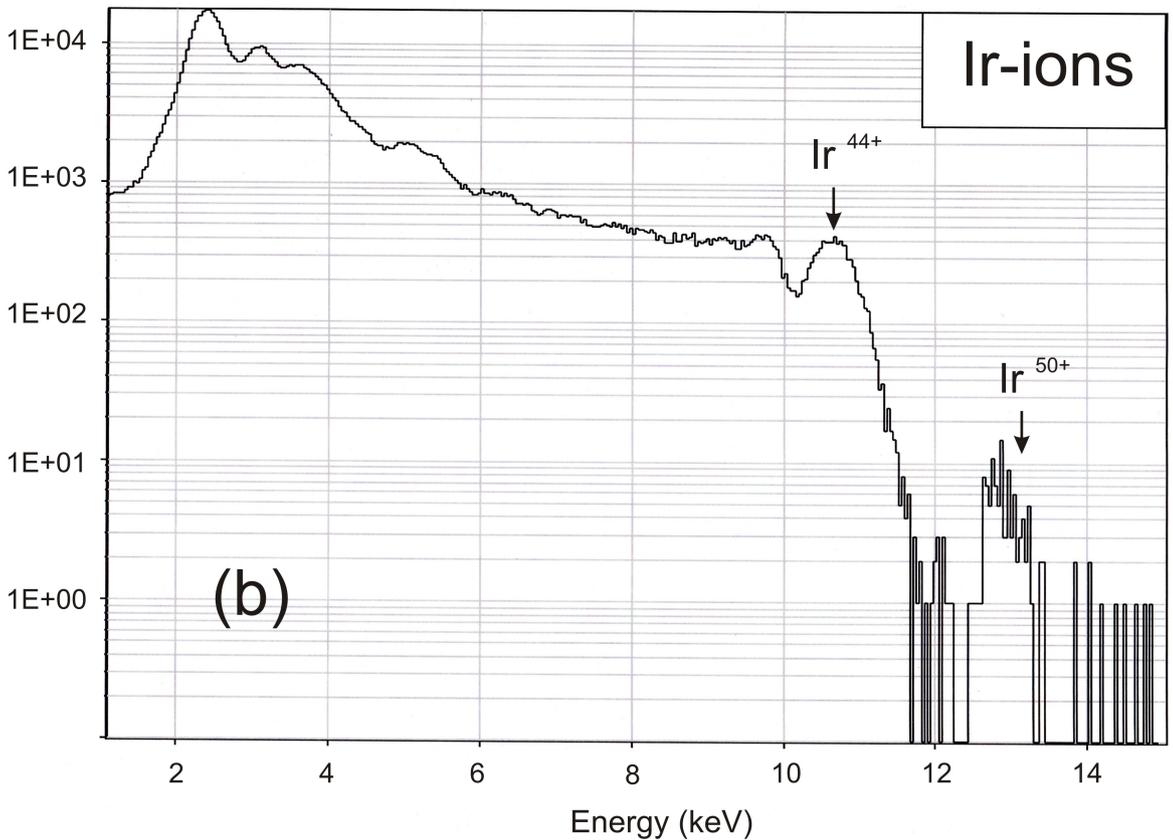

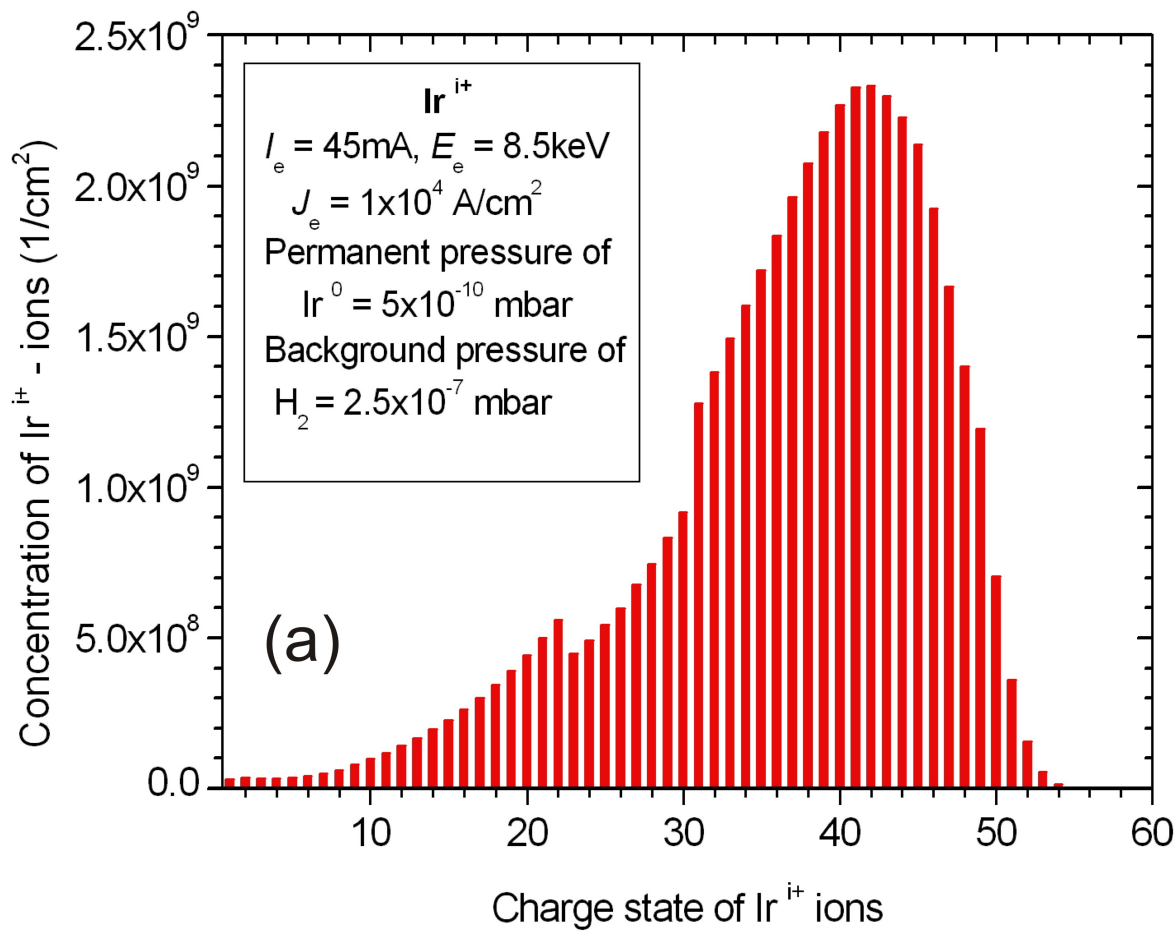
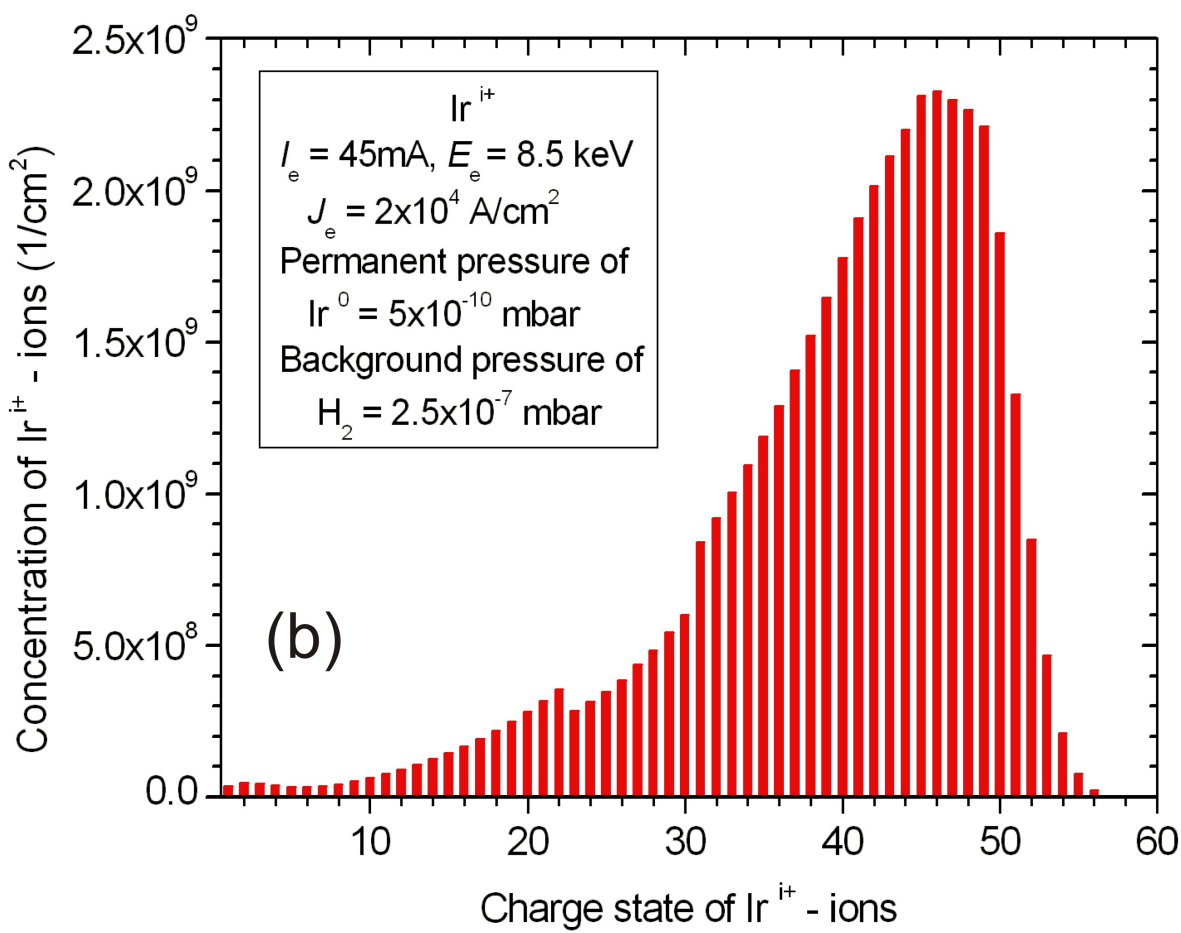